\documentclass[pre,superscriptaddress,showkeys,showpacs,twocolumn,tightenlines]{revtex4-1}
\usepackage{graphicx}
\usepackage{latexsym}
\usepackage{amsmath}
\begin {document}
\title {Stretched exponentials and tensionless glass  in the plaquette Ising model}
\author{Adam Lipowski}
\affiliation{Faculty of Physics, Adam Mickiewicz University, 61-614 Pozna\'{n}, Poland}
\begin {abstract}
Using Monte Carlo simulations we show that the autocorrelation function $C(t)$ in the $d=3$ Ising model with a plaquette interaction has a stretched-exponential decay in a supercooled liquid phase. Such a decay characterizes also some ground-state probability distributions obtained from the numerically exact counting of up to $10^{450}$ configurations. A related model with a strongly degenerate ground state but lacking glassy features does not exhibit such a decay. Althoug the stretched exponential decay of $C(t)$ in the three-dimensional supercooled liquid  is inconsistent with the droplet model, its modification that considers tensionless droplets might explain such a decay. An indication that tensionless  droplets might play some role comes from the  analysis of low-temperature domains that compose the glassy state. It shows that the energy of a domain of size $l$ scales  as $l^{1.15}$, hence these domains are indeed tensionless .
\end{abstract}
\pacs{} \keywords{}

\maketitle
\section{Introduction}
Assuming that liquids constitute a homogeneous and continuum medium, one can explain many of its properties such as viscosity, diffusion or chemical reaction rates. However, cooling below melting point causes drastic changes in the dynamics and a homogeneous approach is no longer legitimate \cite{ediger}. One of the landmarks of this supercooled regime is a slower than exponential decay of autocorrelation functions. Although this decay is often fitted with the so-called stretched exponentials $\exp[-(t/\tau)^{\beta}]$, it is sometimes questioned as being merely a phenomenological fit to experimental data without any clear microscopic mechanism \cite{gotze}. 

It is becoming commonly accepted that supercooled liquids are dominated by dynamically generated heterogeneities. They have different sizes and lifetimes and their dynamics is thus very complex. It would be desirable to relate this dynamics with some kind of an Ising model the dynamics of which is relatively understood. Particularly interesting in this  context might be the droplet model, which predicts  in some cases the stretched-exponential decay of autocorrelation functions \cite {huse}. According to the droplet model, however, such a decay should hold only for low-dimensional Ising systems, and in the most interesting three-dimensional case, an exponential decay is expected. 

A promising statistical mechanics approach to glassy systems refers to lattice models.
Various models of glasses, including kinetically constrained or spin-facilitated ones have already been examined \cite{ritort}. Although they exhibit an interesting slow dynamics, their thermodynamics is very often trivial. A more comprehensive description of glassy systems, that would possibly encompass also superliquid and crystal phases, might be thus sought among Ising models. 

In the present paper we examine a certain three-dimensional Ising model for which it was already shown that it exhibits some glassy features.  We show that  autocorrelation functions of this model have the expected  in the supercooled liquid phase stretched exponential decay. We also find the stretched-exponential decay in some ground-state probability distributions of this model. It is not clear to us whether such a decay of ground-state probability distributions is related with finite-temperature decay of autocorrelation functions, but we noticed that in a related model which is lacking the glassy features, these probability distributions have the ordinary exponential decay. 
Finally, in a somewhat speculative way, we argue, that stretched-exponential decay of autocorrelation functions in three-dimensional systems can be reconciled with the droplet model  provided that these droplets are tensionless. Such an assumption finds some support in the anaysis of the low-temperature  domains in the glassy phase.

\section{Model}
In the present paper, we examine the Ising model with plaquette interaction, that is defined using the following Hamiltonian
\begin{equation}
H = - \sum_{(i,j,k,l)} S_iS_jS_kS_l, \ \ \ S_i=\pm 1, \ \  i=1,2,\ldots, L^d 
\label{hamilton}	
\end{equation}
where summation is over elementary plaquettes of the $d$-dimensional Cartesian lattice of the linear size $L$ with periodic boundary conditions. For $d=3$, model (\ref{hamilton})  shares a number of properties with glassy systems. In particular, it exhibits a strong metastability  \cite{lip97,bray} and a very  slow (perhaps logarithmically slow) coarsening dynamics \cite{lipdesespriu}. Moreover, aging \cite{bray} and cooling-rate effects \cite{lipdes} are consistent with expectations for ordinary glasses. 
Let us also notice that even the $d=2$ version of the model, despite trivial thermodynamics, exhibits an interesting glassy behaviour \cite{buhot,jack2005}.

\section{Spin-spin correlation functions}

\begin{figure}
\includegraphics[width=9cm]{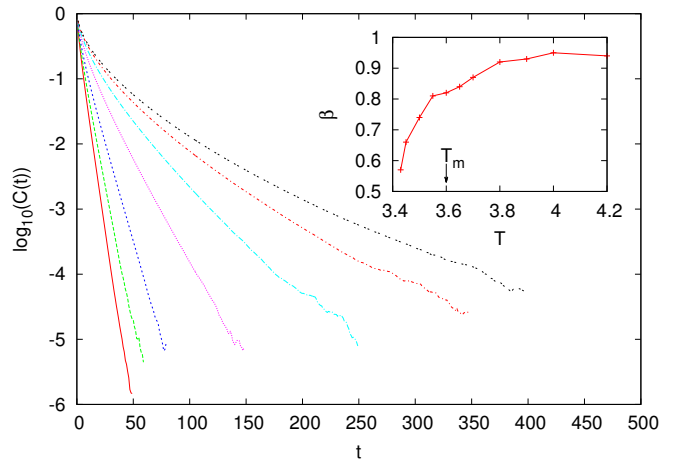}  \vspace{-0.7cm} 
\caption{(Color online) The time dependence of the spin-spin autocorrelation function $C(t)$ for (from above) $T=3.43$, 3.45, 3.5, 3.6, 3.8, 4.0, and 4.2.  At each temperature, simulations start from a random initial configuration, which is relaxed for a certain time ($\sim 10^3$), and then measurements are made during $t\sim 10^7$ ($L=50$). A unit of time corresponds to a single on average update per spin. The inset shows the exponent $\beta$ obained from the fit to the stretched exponential ($a\exp {-(t/\tau)^{\beta}}$). The arrow indicates location of the first-order melting transition in the model.
\label{cspin}
}
\end{figure}

To examine the dynamics  of model (\ref{hamilton}) in the supercooled liquid phase, we calculated the spin-spin autocorrelation function that is defined as  
\begin{equation}
\label{corel}
C(t) =  \frac{1}{L^3}\sum_{i} \langle S_i(0)S_i(t)\rangle
\end{equation}
Using a standard Metropolis dynamics, we simulated the  model for the $d=3$ case at several temperatures and measured $C(t)$.
The results of these simulations are shown in Fig.\ref{cspin}. Fitting our data to the function $a\exp {[-(t/\tau)^{\beta}]}$, we obtained the exponent $\beta$, and its temperature dependence is shown in the inset. Let us notice that earlier studies of model (\ref{hamilton}) locate the glassy transition close to $T=T_g\approx 3.4$  and the first-order melting transition (located on the comparison of free energies and simulations with nonhomogeneous initial conditions) around $T=T_m\approx 3.6$ \cite{lip97,bray,devatol}. One can notice that for $T<T_m$ an appreciable departure from the exponential ($\beta=1$) decay is seen. 
It is not clear to us whether $\beta$ becomes smaller than 1 precisely at $T_m$ or at a somewhat larger value, as our data might suggest. 
Studies of long-time evolution of glassy systems using molecular dynamics simulations of realistic systems are computationally very demanding, but in a model with a controlled frustration, Shintani reported a similar temperature dependence of $\beta$ \cite{shintani}. A stretched-exponential behaviour of the energy autocorrelation function with $\beta$ decreasing upon approaching $T_g$ has already been reported for model (\ref{hamilton}) \cite{bray}. However, energy is  a global variable and it is not clear whether such a quantity correctly probes the heterogenoeus dynamics of supercooled liquids. There are also some reports of a stretched exponential behaviour in other Ising-like lattice models  with glassy features \cite{cavagna} but it might be a consequence of a reduced ($d=2$) dimensionality, which according to the droplet model \cite{huse} might imply such a decay of $C(t)$.  

\section{Stretched exponentials at $T=0$}
\begin{figure}
\begin{center}
\includegraphics[width=6cm]{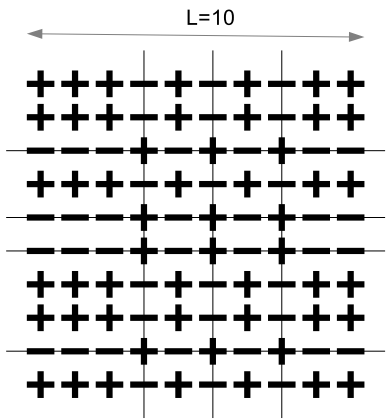}  \vspace{-1.0cm} 
\end{center}
\caption{ An example of a ground-state configuration in the $d=2$ case obtained from the ferromagnetic state (all +) by flipping $k_1=4$ horizontal and $k_2=3$ vertical rows of spins.
\label{config}}
\end{figure}

As we show below, stretched exponentials appear in model (\ref{hamilton}) also in a much different context. Namely, they characterize some ground-state probability distributions. A calculation of these distributions is possible because  a strongly degenerate ground state has in fact a relatively simple structure. For example, for $d=2$ all its configurations can be obtained from a reference configuration (e.g., all spins +) by flipping vertical and horizontal rows of spins (Fig.\ref{config}), which  leads to its $2^{2L-1}$ degeneracy \cite{jack2005}.  

To illustrate the calculations, let us examine the susceptibility-like variable  $\chi_2 =\frac{1}{L^2}\sum_{i,j} S_iS_j=\frac{1}{L^2}(\sum_{i} S_i)^2$. One can notice that to calculate $\chi_2$ precise distribution of these flipped rows is not needed, and it is sufficient to know only their numbers $k_1$ and $k_2$. Indeed, flipping $k_1$ horizontal rows, we reduce the number of $+$ spins to $N_{+}=L^2-k_1L$. The subsequent flip of $k_2$ vertical rows leads to $N_{+}=L^2-(k_1+k_2)L+2k_1k_2$. 
Using $\sum_{i} S_i=2N_{+}-L^2$, we obtain that for $(k_1,k_2)$ configurations $\chi_2=\frac{1}{L^2}[L^2-2(k_1+k_2)L+4k_1k_2]^2$.  Of course, calculating the probability distribution of $\chi_2$, one should take into account the multiplicity factor equal to $\binom {L-1} {k_1}  \binom {L-1} {k_2}$.

The above considerations can be easily generalized to the $d=3$ version. In this case, ground-state configurations can be obtained from the reference configuration by flipping entire two-dimensional planes. To characterize a given ground-state configuration, we now need three numbers $(k_1,k_2,k_3)$, for which we obtain
\begin{eqnarray} \chi_3 & = & \frac{1}{L^3}[L^3-2(k_1+k_2+k_3)L^2 \nonumber \\ & & + 4(k_1k_2+k_1k_3+k_2k_3)L-8k_1k_2k_3]^2. 
\label{susc3}
\end{eqnarray}
The corresponding multiplicity factor equals  $\binom {L-1} {k_1}  \binom {L-1} {k_2} \binom {L-1} {k_3}$.

For a further analysis of the probability distribution, we resort to numerical caclulations. For $d=3$ and a given system size $L$,  we generate all $(L-1)^3$ triples $(k_1,k_2,k_3)$,  and using Eq.(\ref{susc3}), we calculate $\chi_3$ and the corresponding multiplicity factor. Collecting the data in some bins, we obtain the required probability distribution $P(\chi_3)$. Let us notice that computational complexity of generation of such triples is rather modest ($\sim L^3$), which allows us to examine large systems ($L\sim 500$) within few seconds of CPU time.

\begin{figure}
\begin{center}
\includegraphics[width=9cm]{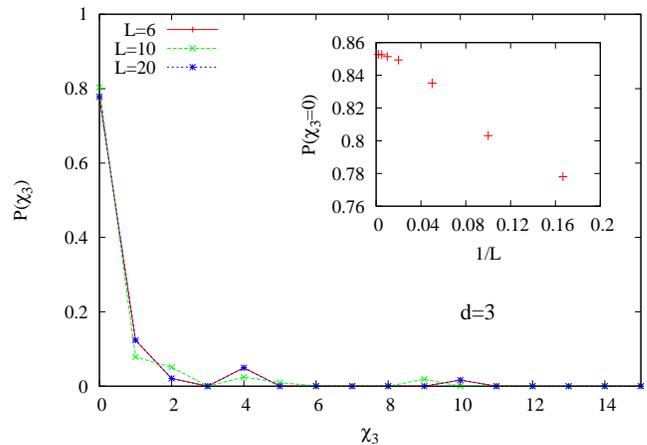}  \vspace{-1.3cm} 
\end{center}
\caption{(Color online) The probability distribution for the susceptibility $\chi_3$. The fact that $\chi_3=0$ is the most typical value of these distributions agrees with the recent Monte Carlo simulations for  lattices of similar size showing that at low temperature the susceptibility drops to zero \cite{des2012}. The inset shows $P(\chi_3=0)$ as a function of $1/L$ for $L=6, 10,\ldots, 500$.
\label{groundd3small}}
\end{figure}

The results presented in Fig.\ref{groundd3small} show that $P(\chi_3)$ has a maximum at $\chi=0$. Let us notice, however, that the average over all ground-state configurations $\langle \chi_3 \rangle=1$. This is a consequence of the flipping symmetry of the Hamiltonian (\ref{hamilton}), which implies that for any $i\neq j$ the corresponding correlation function $\langle S_iS_j\rangle$ vanishes \cite{lip97}. The only nonvanishing contribution to susceptibility comes from the case $i=j$ and that implies $\langle \chi_3 \rangle=1$.

The above symmetry arguments are valid also at finite (and sufficiently high) temperature and recent Johnston's Monte Carlo simulations  report indeed $\langle \chi_3 \rangle \approx 1$ \cite{des2012}. These simulations also show that at low temperature the susceptibility drops almost to zero. Our results (Fig.\ref{groundd3small}) shed some light on such a finding: Monte Carlo simulations at low temperature select randomly one of the ground states (toward which the system slowly evolves) and since $\chi_3=0$ is the most probable value in $P(\chi_3)$, this is the value that is typically measured in Monte Carlo simulations. Let us notice that due to the strong degeneracy of the ground state, it is difficult to find the order parameter that would distinguish low and high temperature phases of the model (\ref{hamilton}). Johnston's result suggests that the susceptibility might serve as such, but $P(\chi=0)$ for increasing $L$ does not converge to unity (inset in Fig.\ref{groundd3small}) and there is a finite (albeit small) probability that simulations will select the ground state with $\chi>0$.

\begin{figure}
\begin{center}
\includegraphics[width=9cm]{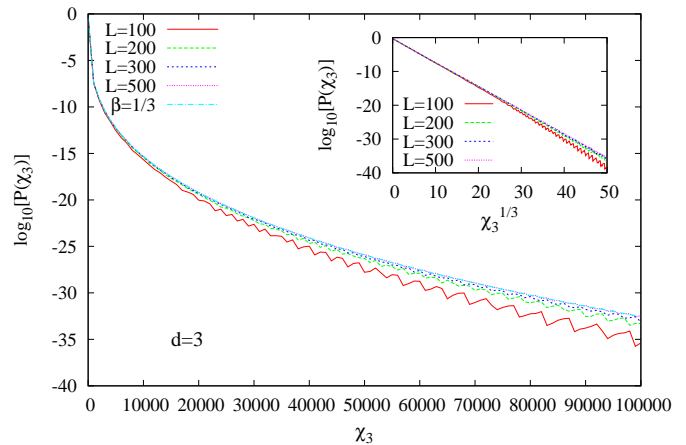}  \vspace{-1.3cm} 
\end{center}
\caption{(Color online) Logarithm of  the probability distribution $P(\chi_3)$ for the three-dimensional version. For $L=500$, the data nearly overlap with the least-square fit with $\beta=1/3$.  The inset confirms the stretched exponential decay with $\beta=1/3$ for nearly 35 decades. 
\label{groundd3}}
\end{figure}

Further analysis of $P(\chi_3)$   (Fig.\ref{groundd3}) shows a slower than exponential decay for large $\chi_3$. Plotting against $\chi_3^{1/3}$ shows an excellent linearity of the data for nearly 35 decades and indicates that asymptotically $P(\chi_3)$ decays as stretched exponential $\exp(-a\chi_3^{1/3})$. Let us also notice that for $L=500$ the obtained probability distribution is based on the (numerically) exact counting of $2^{3 \cdot 500 -2}\sim 10^{450}$ ground-state configurations.

\begin{figure}
\begin{center}
\includegraphics[width=9cm]{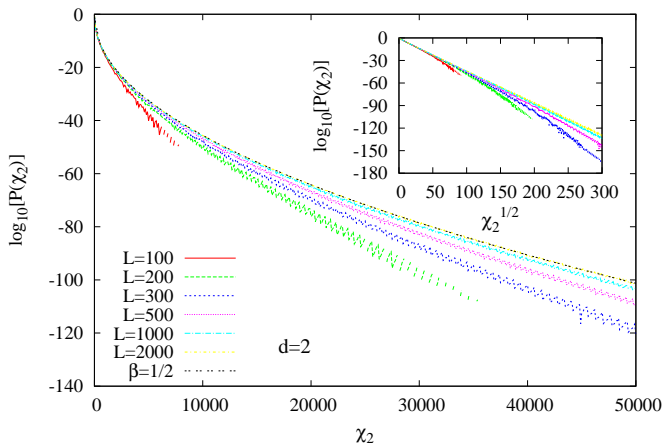}  \vspace{-1.3cm} 
\end{center}
\caption{(Color online) Logarithm of  the probability distribution $P(\chi_2)$ for the two-dimensional version. For $L=2000$, the data nearly overlap with the least-square fit with $\beta=1/2$.  The inset confirms the stretched exponential decay with $\beta=1/2$ for nearly 120 decades. 
\label{groundd2}}
\end{figure}

Although the $d=2$ version of the model (\ref{hamilton}) has a trivial thermodynamic behaviour, it exhibits an interesting glassy-like dynamic behaviour \cite{jack2005}. Calculating the distribution $P(\chi_2)$, we also find the stretched exponential decay but with the exponent $\beta=1/2$ (Fig.\ref{groundd2}).

It is not clear to us whether stretched exponential (static) distributions that we found at the ground state of model (\ref{hamilton}) are related  at all with its finite-temperature dynamic behaviour. However, a curious example comes from a more general version of model (\ref{hamilton})  known as the  gonihedric model. This model, introduced in the context of the discretized string theory \cite{savvidy}, is described by the following Hamiltonian
\begin{equation}
H = -2\kappa \sum_{i,j} S_iS_j +\frac{\kappa}{2} \sideset{}{'}\sum_{i,j} S_iS_j -\frac{1-\kappa}{2} \sum_{(i,j,k,l)} S_iS_jS_kS_l 
\label{hamilton-goni}	
\end{equation}
where the first and the second summations are over the nearest and the next-nearest neighbours, respectively. For $\kappa=0$, the model reduces to the plaquette model (\ref{hamilton}). A quite different thermodynamic and dynamic behaviour is reported for $\kappa\neq 0$ \cite{cirillo,lipdesespriu}. In such a case, there is no metastability  upon temperature changes and the dynamics does not exhibit glassy features. Also the flipping symmetry of the model (\ref{hamilton-goni}) is lower than in the case of $\kappa=0$. In particular, flipped planes or rows of spins cannot cross \cite{cirillo}. This simplifies the calculations  we made for the plaquette model since now only one of the numbers $k_1$, $k_2$ or $k_3$ might be nonzero. In the $d=3$ version of the gonihedric model for a configuration with $k$ planes flipped, we thus obtain 

\begin{equation} \chi_{3\kappa}=\frac{1}{L^3}[L^3-2kL^2]^2 = L[L-2k]^2. 
\label{susc3-goni}
\end{equation}
and the multiplicity factor  being equal to $3\binom{L}{k}$ (with prefactor 3 corresponding to the number of directions of flipped planes). Using the above equation, one obtains
\begin{equation}
P(\chi_{3\kappa})=\frac{6}{3\cdot 2^{L}} \binom{L}{\frac{1}{2}(L-\sqrt{\chi_{3\kappa}/L})}
\label{prob-chi}
\end{equation}
where $3\cdot 2^L$ is the degeneracy of the ground state.
Using the identity $\binom{L}{k}=\frac{\binom{L}{L/2}(L/2)!(L/2)!}{(L-k)!k!}$ and the asymptotic form $\binom{L}{L/2}\approx\frac{2^L}{\sqrt{\pi L}}$, after some calculations one obtains that for $1\ll \sqrt{\chi_{3\kappa}/L}\ll L$
\begin{equation}
P(\chi_{3\kappa})\approx \frac{2\exp{(\frac{-\chi_{3\kappa}}{2L^2})}}{\sqrt{\pi L}}
\label{prob-asymptot}
\end{equation}

\begin{figure}
\begin{center}
\includegraphics[width=8cm]{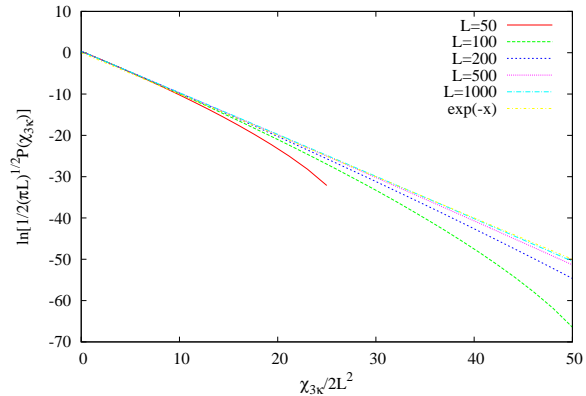}  \vspace{-0.7cm} 
\end{center}
\caption{(Color online) Rescaled logarithm of  the probability distribution $P(\chi_{3\kappa})$ for the three-dimensional version of the gonihedric model (\ref{hamilton-goni}) for $\kappa \neq 0$. For $L=1000$, the data nearly overlap with the straight line corresponding to $\exp{(-x)}$ confirming thus the asymptotic estimation (\ref{prob-asymptot}). 
\label{groundk}}
\end{figure}

Our numerical data are in a very good agreement with the above estimation (Fig.\ref{groundk}). Let us notice that $P(\chi_{3\kappa})$ has an exponential decay that in the thermodynamic limit $L\rightarrow \infty$ flattens and $P(\chi_{3\kappa})\rightarrow 0$. Thus the ground-state distribution in the gonihedric model for $\kappa \neq 0$ has a much different form than in the $\kappa=0$ case. Perhaps it is only a coincidence that in the latter case, where the model exhibits glassy behaviour, the distribution $P(\chi_3)$ has a stretched-exponential behaviour. But one cannot exclude a more profound relation between zero-temperature statics and finite-temperature dynamics of model (\ref{hamilton}). 

\section{Tensionless droplet model}
\label{section-droplet}
Now we would like to return to the problem of the decay of $C(t)$ in the supercooled regime.  It would be desirable to explain such a dynamics in terms of some kind of the droplet model \cite{huse}. The droplet model most likely provides a qualitatively correct description of the Ising dynamics \cite {ito,stauffer} but it predicts the exponential decay of $C(t)$ for $d=3$ systems. It seems to us, however, that trying to use the droplet model to explain the dynamics of model (\ref{hamilton}),  we might have to modify some of its assumptions. Indeed, 
the excess energy $e_l$ of a droplet  of linear size $l$, which for an ordinary Ising model is proportional to its surface ($e_l \sim l^{d-1}$), in model (\ref{hamilton}) might scale as $e_l \sim l^{d-2}$ \cite{lip97,cirillo}.  Actually, droplets with energy proportional to $l^{d-1}$ are also possible in model (\ref{hamilton}) but in our opinion the long-term dynamics might be under the influence of mainly the (low-energy) tensionless droplets. Assuming $e_l\sim l^{d-1}$  and repeating Lifshitz reasoning, one  obtains that the deterministic motion of a droplet satisfies $dl/dt\sim l^{-2}$ and the resulting lifetime of a domain of the initial size $l$ thus scales as $l^3$. Consequently, the form of the Boltzmann factor leads to  the estimation $C(t)\sim \exp {[-(t/\tau)^{(d-2)/3}]}$. Thus,  tensionless domains in model (\ref{hamilton}) might imply stretched exponential behaviour  in three-dimensional systems.

The dynamics of model (\ref{hamilton}) in the supercooled liquid phase is very complex and most likely the entire spectrum of droplets is present. Droplets with energies scaling as $l^{d-2}$ and $l^{d-1}$ are only  limiting cases and perhaps a more adequate description could be obtained asssuming $e_l\sim l^{\phi}$, where $d-2 < \phi < d-1$.  In such a case, we are   lead to $C(t)\sim \exp {[-(t/\tau)^{\phi/(d+1-\phi)}]}$.  Although it would be very desirable, calculation of  surface tension of supercooled droplets in model~(\ref{hamilton}), that would verify our speculative arguments,  is likely to be very difficult. In the last section, however, we provide some numerical results showing that at least the glassy phase might be considered as tensionless.

\section{ Structure of glassy phase}
\begin{figure}
\begin{center}
\includegraphics[width=\columnwidth]{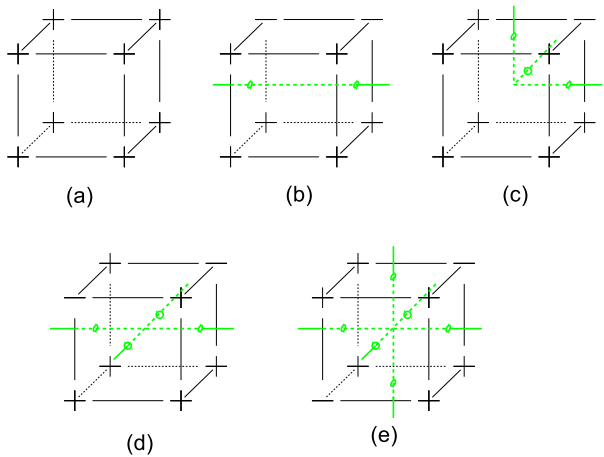}  \vspace{-0.7cm} 
\end{center}
\caption{(Color online) An example of an elementary cube with 0 (a), 2 (b), 3 (c), 4 (d), and 6 (e) unsatisfied plaquettes. Normals to unsatisifed plaquettes are shown in green.
\label{cubes}}
\end{figure}

To get additional insigt into the behaviour of model (\ref{hamilton}) in the $d=3$ case we analysed distributions of unsatisifed plaquettes. In particular we calculated concentrations $c_i$ of elementary cubes that have $i$ unsatisfied plaquettes (see Fig.\ref{cubes}). Elementary analysis shows that there is no configuration of spin variables that would lead to cubes with 1 or 5 unsatisifed plaquettes. 

\begin{figure}
\begin{center}
\includegraphics[width=\columnwidth]{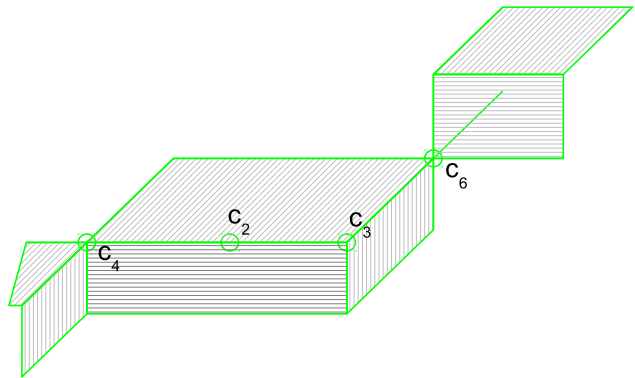}  \vspace{-0.7cm} 
\end{center}
\caption{(Color online) An example of a structure with 2-, 3-, 4-, and 6-unsatisfied cubes (green lines join unsatisifed plaquettes). One can notice that the excess energy is localized only in  edges and corners, and thus scales linearly with the size of the excitation. 
\label{cube-types}}
\end{figure}

Although some simple tensionless structures  can be constructed "by hand" (Fig.\ref{cube-types}), it is not obvious that such structures actually form as a result of the dynamic evolution of the model. To approach such a problem we present zero-temperature configurations obtained using the linear cooling  $T(t)=T_0-r t$, where $r$ is the cooling rate. The value of the initial temperature $T_0$ is unimportant, as long as it is inside the liquid phase (we used $T_0=3.5$). Resulting structures (Fig.\ref{config}) are void of cubes with 6 unsatisfied plaquettes. For the lowest examined cooling rate $r=10^{-7}$ even cubes with 4 unsatisfied plaquettes are almost expelled from the system. 

\begin{figure}
\begin{center}
\includegraphics[width=\columnwidth]{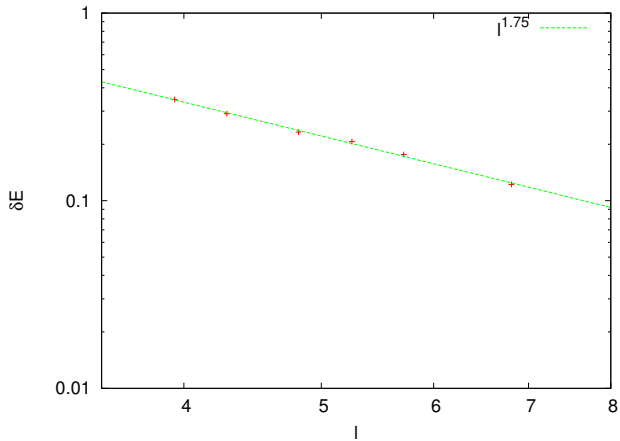}  \vspace{-0.7cm} 
\end{center}
\caption{(Color online) The excess energy $\delta E$ as a function of characteristic length $l$. Approximate fit $\delta E \sim l^{1.85}$ indicate that energy of a domain of size $l$ scales as $l^{1.15}$.    Presented data correspond to zero-temperature averages obtained during linear cooling with the cooling rate $10^{-5}<r<10^{-2}$.  
\label{ener-l}}
\end{figure}

Especially in the latter case one can see that slowely cooled glass has a  cuboid-like structure with (excess) energy, accumulated almost entirely in edges. We expect that for such a structure energy  scales slower than the surface of these domains and most likely it is proportional  to their linear size \cite{surf-tension}. More quantitative confirmation of  the tensionless character of the glassy phase in model (\ref{hamilton}) is obtained from the relation of the zero-temperature excess energy $\delta E = \langle H \rangle/N-E_0$, where $E_0=-3$ is the ground state energy of model (\ref{hamilton}), and the average linear size of domains, defined as an average length of a linear segment \cite{linear}. We calculated  $\delta E$ and $l$ at the end of a cooling process with $10^{-5}<r<10^{-2}$ and the resulting data (averaged over $\sim 100$ independent runs) are shown in Fig.\ref{ener-l}.  Assuming that a final configuration is composed of domains of size $l$ whose energy scales as $l^{\phi}$, we easily arrive at the relation $\delta E \sim l^{\phi-3}$. From the approximate linearity of numerical data (Fig.\ref{ener-l}) we conclude that $\phi\approx 1.15$.  While we cannot definitely resolve whether $\phi$ equals to unity or it gets a certain nontrivial value greater than unity,  our data clearly show that $\phi<2$. It means that energy increases with $l$ slower than the surface ($\sim l^2$) and thus low-temperature domains are tensionless.

\begin{figure}
\begin{center}
\includegraphics[width=\columnwidth]{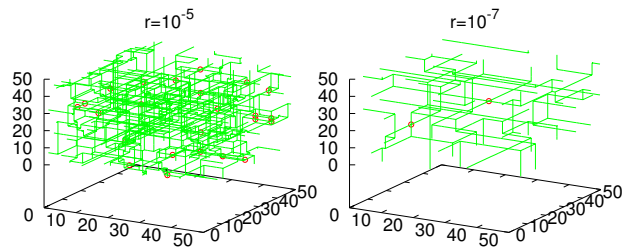}  \vspace{-0.7cm} 
\end{center}
\caption{(Color online) Structure of $T=0$ glass obtained for different cooling rate $r$ and $L=50$. All axes are scaled with lattice sites.  Lines join the unsatisfied plaquettes and small circles indicates elementary cubes with four unsatisifed plaquettes. The above cooling rates are sufficiently small to entirely remove cubes with six unsatisfied plaquettes.   
\label{glass-config}}
\end{figure}

\begin{figure}
\begin{center}
\includegraphics[width=\columnwidth]{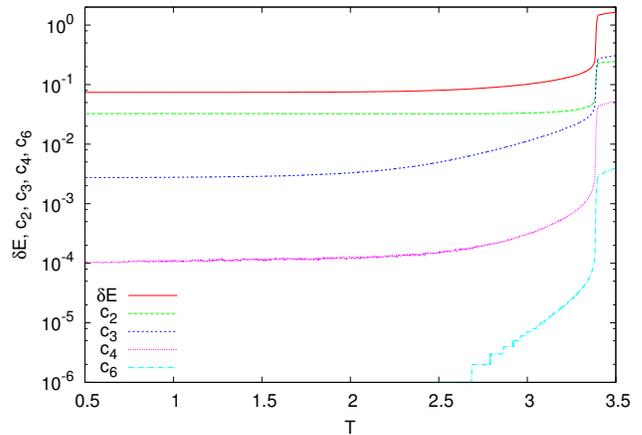}  \vspace{-0.7cm} 
\end{center}
\caption{(Color online) The energy difference $\delta E=E-E_0$ ($E_0=-3$) and concentrations of  2-, 3-, 4-, and 6-unsatisfied cubes as a function of temperature. Simulations were done for $L=100$, $r= 10^{-6}$  and average over $10$ runs was taken.   
\label{conct}}
\end{figure}

\begin{figure}
\begin{center}
\includegraphics[width=\columnwidth]{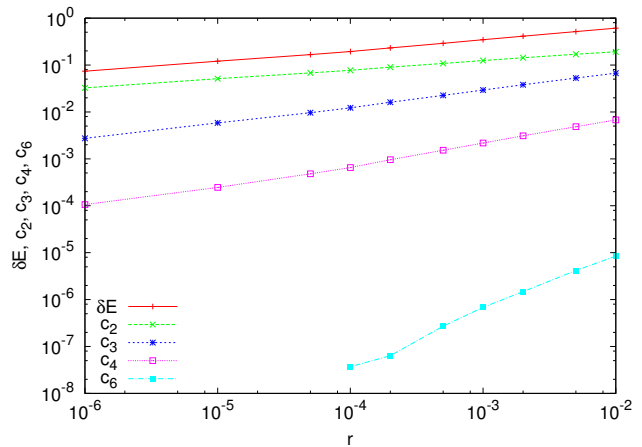}  \vspace{-0.7cm} 
\end{center}
\caption{(Color online) Zero-temperature energy difference $\delta E$ and concentrations of 2-, 3-, 4-, and 6-unsatisfied cubes as a function of cooling rate $r$. Simulations were done for $L=100$ and average over at least $10$ runs was taken.   
\label{conct0}}
\end{figure}

We also examined the temperature dependence of concentrations $c_2$, $c_3$, $c_4$, $c_6$ and of $\delta E$ (Fig.\ref{conct}). All quantities drop at the glassy transition $T_g$ and the largest decrease is seen for (energy-rich) $c_6$ and $c_4$.  Repeating the simulations for several cooling rates  we can examine $r$-dependence of $c_2$, $c_3$, $c_4$, $c_6$, and $\delta E$ at zero temperature (Fig.\ref{conct0}). As expected, the fastest decay for decreasing $r$ exhibit $c_6$ that measures the concentration of high-energy cubes. The slowest decay is seen for  $c_2$ and our data suggest a power-law decay $c_2(r)\sim r^{0.2}$. Let us notice that assuming cuboid-like structure of glassy phase with a characteristic length $l$ we obtain a relation $l \sim c_2^{-1/2}$ and thus $l \sim r^{-0.1}$. The exponent 0.1 is however quite small and we cannot exclude that asymptotically $c_2$ will have slower than a power-law decay. As a matter of fact there are some phenomenological arguments that in glassy systems the characteristic length scale should diverge logaritmically with the cooling rate \cite{shore}. For very slow cooling cubes with 2 unsatisfied plaquettes make dominant contributions to the excess energy $\delta E$ and asymptotically $\delta E$ and $c_2$ are likely to have the same r-dependence. 

\section{conclusions} 
Numerically exact calculations of some ground state probability distribution for model (\ref{hamilton}) show   stretched-exponential decay spaning over more than 30 decades. We hope that such a decay will be confirmed using more rigorous analytical  arguments. 
The stretched-exponential decay is often merely a phenomenological fit to experimental or numerical data. Thus model (\ref{hamilton}) might be of the simplest glassy systems where some insigth into the microscopic mechanism leading to  such decay could be obtained. 

Perhaps the most interesting of our results is at the same time the most speculative one. Suggestion that tensionless droplets might be responsible for the stretched-exponential decay of autocorrelation functions in the supercooled liquid phase find, however, some support in the analysis of the zero-temperature structure of glassy phase, but more convincing arguments, based perhaps on more detailed analysis of the dynamics  of supercooled liquid would be certainly desirable. If confirmed, the remaining question will be whether the proposed mechanism of stretched-exponenital decay has any relevance for real glasses.

\begin{acknowledgments}I thank Des Johnston for sending me his paper prior to the publication. 
\end{acknowledgments}


\begin{thebibliography}{}
\bibitem{linear}  Linear segment is defined as a connected sequence of  cubes with 2 unsatisfied plaquettes ((b) on Fig.\ref{cubes}). Such a sequence must  terminate with cubes having 3, 4 or 6 unsatisfied plaquettes. We assume that the average length of such a segment sets a characterisitc length in the system (corresponding approximately to a  linear size of domains).
\end{thebibliography}
\end {document}